\documentclass[aps,prl,11pt,footinbib, showpacs,showkeys]{revtex4}
\usepackage{graphicx}
\usepackage{epsfig}
\usepackage{dcolumn}
\usepackage{bm}
\usepackage{amsmath}
\usepackage{amsbsy}
\usepackage{amssymb}
\usepackage{color}
\usepackage{verbatim}
\begin{document}

\title{Twisted-light-induced intersubband transitions in quantum wells at normal incidence}

\author{B.\ Sbierski}
\affiliation{$^1$Physik-Department, 
Technische Universit\"at M\"unchen, 
James-Franck-Str.\ 1, 85748 Garching, Germany}

\author{G.\ F.\ Quinteiro}
\author{P.\ I.\ Tamborenea}
\affiliation{$^2$Departamento de F\'isica and IFIBA, FCEN, 
Universidad de Buenos Aires, 
Ciudad Universitaria, Pab.\ I, 
Ciudad de Buenos Aires, Argentina}



\begin{abstract}
We examine theoretically the intersubband transitions induced by laser beams 
of light with orbital angular momentum (twisted light) in semiconductor 
quantum wells at normal incidence.
These transitions become possible in the absence of gratings thanks to 
the fact that collimated laser beams present a component of the light's
electric field in the propagation direction.
We derive the matrix elements of the light-matter interaction for
a Bessel-type twisted-light beam represented by its vector potential in the
paraxial approximation.
Then, we consider the dynamics of photo-excited electrons making intersubband 
transitions between the first and second subbands of a standard semiconductor 
quantum well.
Finally, we analyze the light-matter matrix elements in order to evaluate 
which transitions are more favorable for given orbital angular momentum of 
the light beam in the case of small semiconductor structures.
\end{abstract}

\maketitle

\section{Introduction}

Quantum wells are a fundamental and well-studied type of man-made semiconductor 
nanostructure \cite{dav}.
They can exhibit remarkable transport phenomena and their optical 
properties have motivated intense basic research and numerous applications.
In particular, intersubband transitions, that is, light-induced transitions
between states of a given energy band belonging to different subbands, have 
been extensively studied and successfully applied to the development of 
infrared lasers and detectors \cite{pai}.
In parallel to these developments in the area of semiconductor nanoscience 
and technology, a new branch of optics developed vigorously in the last 
twenty years, namely, the study of phase-structured light \cite{and}.
The generation and applications of {\it twisted light} (TL), or light beams
carrying {\it orbital} angular momentum, gained great attention after the
seminal work of Allen {\it et al.} \cite{all-bei-spr}
The interaction of TL with mesoscopic particles 
(optical tweezers) \cite{bar-tab,all,fri-nie-hek-rub},
atoms and molecules \cite{dav-and-bab,ara-ver-cla},
and Bose-Einstein condensates \cite{and-ryu-cla,sim-nyg-hu}
has been studied recently.
Other issues, like the theoretical description of cavity-QED 
of TL \cite{ala-bab} and the use of TL for potential 
applications in quantum information processing \cite{mut-str} have also 
been addressed \cite{all-bar-pad}.

The interaction of TL with solid-state systems thus appears 
as a promising field of research and technology.
A number of basic situations have recently been investigated 
theoretically \cite{qui-tam-09a,qui-tam-09b,qui-ber,qui-tam-10,
qui-luc-tam,qui,wat-mos-ber,kok-ber}, but fewer experimental studies
have been reported thus far \cite{uen-tod-ada,cla-mcc-dre}.
In this article we initiate the study of intersubband transitions 
induced by TL beams in semiconductor quantum wells.
Besides considering the intersubband excitation with TL, 
we focus on a special geometry that has not received much attention 
thus far:
By taking advantage of the fact that collimated laser beams posses a 
component of the electric field along the propagation direction,
we propose the excitation of intersubband transitions at normal incidence
without using gratings.
Thus, collimated beams of TL bring about two nonstandard ingredients 
to intersubband transitions: 
new selection rules based on the conservation of the spin and {\it orbital}
angular momentum, and the possibility of excitation at normal incidence.


\section{Twisted light}

Let us consider a TL beam with a radial dependence of the Bessel type and
circular polarization $\sigma=\pm 1$.
In the Coulomb gauge ($\nabla\cdot\mathbf{A}=0$) its vector potential
in the paraxial approximation is given by \cite{vol-etal,jau}
\begin{equation}
    \mathbf{A}(\textbf{r},t)
= A_0 \, e^{i(q_z z- \omega t)}
    \, \left[ \boldsymbol{\epsilon}_{\sigma}
              J_{\ell}(q_r r) e^{i \ell \phi} - i \sigma \, \mathbf{e}_{z}
              \frac{q_r}{q_z} J_{\ell + \sigma}(q_r r)
              e^{i (\ell + \sigma) \phi} 
       \right] + c.c. \,,
\label{Eq_A}
\end{equation}
with $\boldsymbol{\epsilon}_{\sigma} = \mathbf{e}_x + i\sigma \mathbf{e}_y$,
Bessel functions $J_{\ell}$, and $q_{r}^{2}+q_{z}^{2}=\omega^{2} / c^{2}$.
The light's orbital angular momentum is given by the integer $\ell$.
When optical transitions can be excited by both the $xy$ and the $z$ 
components of the light's electric field, the $z$-component of the vector 
potential is normally neglected since usually $q_r \ll q_z$.
On the other hand, there are experimental situations where the component 
of the field in the plane of the absorbing material ($xy$-plane) cannot 
induce quantum transitions due to symmetry restrictions while the 
$z$-component can.
This is the case of intersubband transitions in quantum-well and quantum-disk
structures at normal incidence, which is the situation we focus on here.
Moreover, advances in optical hyperlenses suggest that highly collimated
beams (with large $q_r$) can be obtained \cite{zha-zhe-li}.
Optical transitions induced by the $xy$-component
of the TL beam in bulk \cite{qui-tam-09a,qui-tam-10},
quantum wells \cite{qui-tam-10}, and quantum dots \cite{qui-tam-09b}
have been studied recently.


\section{Semiconductor quantum well}

In order to model the semiconductor quantum well in a convenient and flexible
way we consider a cylindrical disk of height $z_0$ and radius $r_0$.
The limit of $r_0 \rightarrow \infty$ describes the infinite 
quasi-two-dimensional quantum well system, and the finite $r_0$ case will 
allow us to do numerical modeling with realistic beam and nanostructure
parameters.
Note that the chosen cylindrical geometry is convenient from a mathematical 
and conceptual point of view, given the cylindrical nature of the 
beam \cite{qui-tam-10}.
We have in mind standard quantum wells made of, for example, GaAlAs \cite{dav}.
With hard-wall confinement, the envelope-function eigenstates of the 
cylindrical disk are given by
\begin{equation}
\psi_{nm\nu}(r) = \mathcal{N}_{m\nu} J_{m}(q_{m\nu}\, r) \, e^{im\phi}
                  \sqrt{\frac{2}{\pi z_0}} \sin\left(\frac{n\pi z}{z_0} \right),
\label{eq:Wavefunction}
\end{equation}
with $q_{m\nu} = x_{m\nu}/r_0$, $x_{m\nu}$ being the $\nu$-th zero of the
Bessel function $J_m$,
$n=1,2,\ldots$ the subband index,
and the normalization constant
\begin{equation}
  \mathcal{N}_{m\nu} = \frac{1}{r_0 J_m'(x_{m\nu})}.
\label{eq:normalization}
\end{equation}
The eigen-energies are
\begin{equation}
\varepsilon_{nm\nu} = \frac{\hbar^2 }{2m_e^{*}}
                      \left(q_{m\nu}^2 + \frac{\pi^{2}n^2}{z_0^2} \right),
\label{eq:eigenenergy}
\end{equation}
where $m_e^*$ is the effective electron mass in the band under consideration.


\section{Light-matter interaction matrix elements}

The intersubband transitions are governed by the matrix elements
of the electron-TL interaction, $h_I$, whose matrix elements
\begin{eqnarray}
\left\langle n'm'\nu'|h_{I}|nm\nu\right\rangle =
-\frac{e}{m_e}
  \left\langle n'm'\nu'|\mathbf{A}\cdot\mathbf{p}|nm\nu \right\rangle
+ \frac{e^{2}}{2m_e^*}
\left\langle n'm'\nu'|\mathbf{A}^{2}|nm\nu \right\rangle
\end{eqnarray}
need to be worked out ($e=-|e|$ and $m_e$ are the electron charge and 
bare mass, respectively).
It can be easily shown that, under normal incidence, the $A_x$ and $A_y$
components of the vector potential do not contribute to the matrix
elements for intersubband processes  ($n' \neq n$).
Intrasubband transitions are stronlgy supressed by the smallness of their 
light-matter matrix elements (between ``in-plane" single-particle states).
Also, considering typical intersubband frequencies of 
$\hslash\omega \simeq 100 \, \mbox{meV}$, 
both intrasubband single-particle excitations and surface plamons are 
completely off-resonance.
Thus, intrasubband effects can be safely ignored here.
For the component $A_z$ we obtain from the linear term of the interaction
\begin{equation}
-\frac{e}{m_e} \left\langle n'm'\nu'|A_{z} p_z |nm \nu \right\rangle =
    \frac{2\pi \sigma e \hbar A_{0}q_{r}}{m_e q_{z}}
    \eta_{m'\nu',m\nu} \, 
    B_{n'n} 
    \left(\delta_{m'-m,\ell+\sigma} \, e^{-i\omega t}
          - \delta_{m'-m,-(\ell+\sigma)} \, e^{i\omega t} 
    \right),
\label{eq:mez}
\end{equation}
where
\begin{equation}
B_{n'n} = \frac{2nn'}{n'^{2}-n^{2}} \delta_{n'+n,\text{odd}},
\end{equation}
\begin{eqnarray}
\eta_{m'\nu',m\nu} = \mathcal{N}_{m'\nu'} \mathcal{N}_{m\nu}
          \int_{0}^{r_0} rdr\, J_{m'}(q_{m'\nu'} r)
                               J_{\ell+\sigma}(q_{r}r)
                               J_{m}(q_{m\nu} r).
\label{eq:xi}
\end{eqnarray}
The factor $B_{n'n}$ has been obtained under the dipole moment approximation 
(applied only in the $z$-direction), $q_z z_0 \ll 1$, which is amply 
justified in our system.
For transitions between subbands 1 and 2, 
we have $B_{21}=\frac{4}{3}=-B_{12}$.
Within the same dipole approximation, the quadratic term is proportional 
to $\delta_{n'n}$ and therefore drops out for intersubband transitions.
The Kronecker deltas that appear in Eq.\ (\ref{eq:mez}) enforce the conservation 
of angular momentum.
Since the electron undergoes an intraband transition, 
there is obviously no change in the angular momentum associated with the 
periodic part of its Bloch wave function. 
As a consequence, the spin angular momentum of the photon ($\sigma$), 
which in interband transitions accounts for the change in the band angular 
momentum of the electron \cite{qui-tam-09a}, appears now in the selection
rules for the envelope function along with the photon's orbital angular
momentum ($\ell$), as seen also in the study of intraband transitions in 
quantum rings \cite{qui-ber, qui-tam-ber}.

The matrix elements of Eq.\ \eqref{eq:mez} involve integrals with 
three Bessel functions
[see Eq.\ \eqref{eq:xi}] which, to the best of our knowledge, do not admit 
an analytical solution and are somewhat numerically demanding.
However, they can be simplified in cases of practical interest where the 
semiconductor structure is small compared to the beam waist.
This wide-beam approximation has the advantage of keeping the interesting
angular dependence of TL while simplifying the radial
dependence which is not essential for our purposes.
Let us consider a typical quantum well with intersubband transition
energy of $\hslash\omega \simeq 100 \, \mbox{meV}$. 
Now, $q_{r}^{2}+q_{z}^{2}=\omega^{2} / c^{2}$ with $q_r \ll q_z$
implies that $q_{r} \ll \omega/c = 8.3 \times 10^5 \, \mbox{m}^{-1}$,
and therefore the first zero of the beam's Bessel function is at least
a few hundred microns from the center of the beam.
Then, for a large quantum disk of a few tens of microns centered
at the beam axis we can approximate the beam's Bessel function
by the first term of its series expansion.
Furthermore, according to Eq.\ \eqref{eq:mez}, only the values of
$m'=m\pm(\ell+\sigma)$ (+ for absorption and - for emission) lead to
non-vanishing matrix elements.
Then
\begin{eqnarray} 
\hspace{-2cm}
\eta_{m\pm(\ell+\sigma),\nu';m\nu} =
                      \mathcal{N}_{m\pm(\ell+\sigma),\nu'}
                      \mathcal{N}_{m,\nu} 
                      \int_{0}^{r_0} r dr
                      J_{m\pm(\ell+\sigma)}(q_{m\pm(\ell+\sigma),\nu'} r) 
                      J_{\ell+\sigma}(q_r r) 
                      J_{m}(q_{m\nu} r) \nonumber \\
\hspace{-1.5cm}
\simeq \frac{\mathcal{N}_{m\pm(\ell+\sigma),\nu'}
             \mathcal{N}_{m,\nu} r_0^2}{\Gamma(\ell+2)}
       \left(\frac{q_r r_0}{2}\right)^{\ell+\sigma}
       \int_{0}^{1}  du 
          J_{m\pm(\ell+\sigma)}(x_{m\pm(\ell+\sigma),\nu'} u)\, 
          J_{m}(x_{m\nu} u)\, 
          u^{\ell+\sigma+1}\,.
\label{eq:widebeam}
\end{eqnarray}


\section{Physical interpretation}
\label{sec:physical_interpretation}

The matrix elements obtained in the previous section permit the calculation
of the different forms of optical response of the quantum well to normally
incident twisted light.
Equations of motion for the single-particle density matrix provide a
general theoretical framework for such calculations, and can be 
straightforwardly obtained, in the free-carrier regime, for example 
from Ref.\ \cite{qui-tam-10}.
Here, we will examine some basic aspects of the dynamics of the 
photoexcited electrons and argue for the detectability of the 
proposed intersubband transitions.

A central quantity to describe the dynamics of the photoexcited electrons
is the population $\rho_{2m\nu}(t)$ of the second subband states with quantum 
numbers $(m,\nu)$.
In the perturbative regime these populations can be obtained through
the equations of motion that we just mentioned \cite{qui-tam-10}, or more 
directly via ordinary time-dependent perturbation theory \cite{lan-lif}.
To second order in the light field and assuming a laser pulse turned on 
at $t=0$ one obtains for the population of the second-subband states
\begin{eqnarray}
\rho_{2m\nu}^{(2)}(t) &=& 
      4 \left( \frac{2\pi e \hbar A_{0} q_{r}}{m_e q_{z}}\right)^2
      \underset{\nu'}{\sum} 
      f_{m-\ell-\sigma,\nu'} \,
      \eta_{m\nu;m-\ell-\sigma,\nu'}^{2} B_{21}^{2} \nonumber \\   
      && \times
      \frac{\sin^2[(\varepsilon_{2m\nu}-
            \varepsilon_{1,m-\ell-\sigma,\nu'}-\hslash\omega)t/2\hslash]}
      {\left(\varepsilon_{2m\nu}-\varepsilon_{1,m-\ell-\sigma,\nu'}-
       \hslash\omega \right)^{2}} \, ,
\label{eq:pop2-1}
\end{eqnarray}
where $f_{m-\ell-\sigma,\nu'}$ is the Fermi distribution that gives the initial
populations of the first-subband states.
From this expression one obtains the Fermi-Golden-Rule linear time dependence 
in the long-time limit, and for short pulses the time-dependence is quadratic.

\begin{figure}[h]
\includegraphics[scale=1]{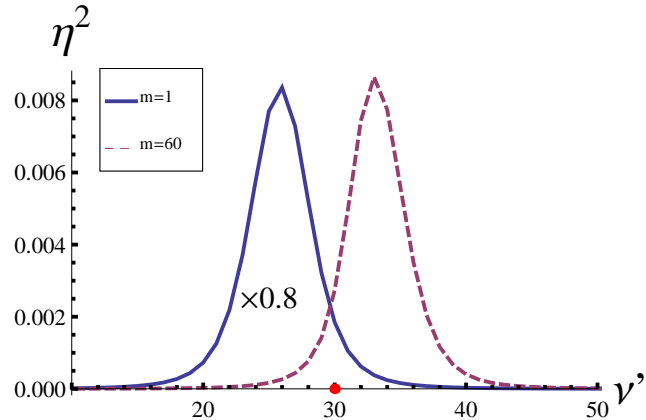}
\caption{Integral $\eta$ squared as a function of $\nu'$ 
with $m'=m+\ell+\sigma$ and for fixed values of $\nu=30$, 
$\ell=10$, and $\sigma=1$, and two different values of $m$. 
Notice the $m$-dependent shift in the peak position.}
\label{fig:1}
\end{figure}

In order to get some insight into the optical excitation and the resulting
population of the second subband in the situation of small nanostructures, 
we have performed numerical calculations of $\eta$ in the case of 
$m'=m+\ell+\sigma$. 
From  Eq.\ \eqref{eq:pop2-1}, one would like to understand what 
contributes to the population of a particular state ($m$, $\nu$) 
of the second subband. 
In Fig.\ \ref{fig:1} we plot $\eta^2$ (we plot $\eta^2$ rather than
$\eta$ since it has the visual advantage of not presenting oscillations 
from positive to negative values), as a function of $\nu'$ for fixed
values of $\nu=30$ and $\ell=10$, and two different values of $m$. 
One immediately recognizes that there are several states $\nu'$ 
of the first subband contributing to the second subband population. 
This is the main difference with the case of vertical 
(one-to-one) transitions induced by plane-waves, and has been 
already discussed in a previous paper \cite{qui-tam-09a}. 
Furthermore, we observe a strong shift in the peak position, 
moving to higher values of $\nu'$ for larger values of $m$.

\begin{figure}
\centerline{\includegraphics[scale=1]{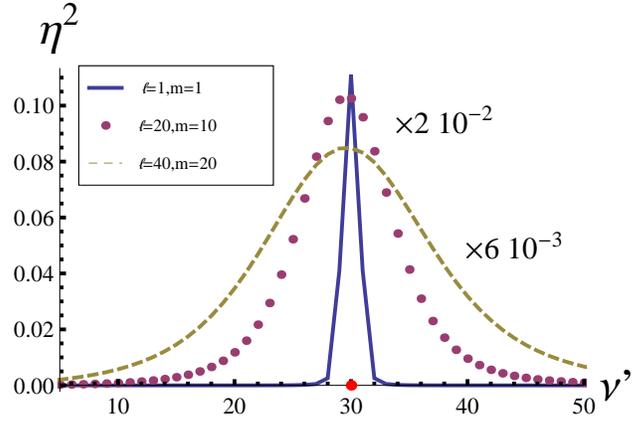}}
\caption{Integral $\eta$ squared as a function of $\nu'$ 
with $m'=m+\ell+\sigma$ for a fixed
value of $\nu=30$ and $\sigma=1$, and three different values 
of $\ell$ and $m$. 
The OAM $\ell$ affects many features: 
i) the width, 
ii) the amplitud and
iii) the shift of the peak position. 
The shift of the peak also depends on the value of $m$, but that  
change is opposite to that of $\ell$: the plot shows how the shifts produced
by $m$ and $\ell$ can compensate each other and produce responses
roughly centered at $\nu=\nu'$.}
\label{fig:2}
\end{figure}

In Fig.\ \ref{fig:2} we plot $\eta^2$ for a fixed
value of $\nu=30$ and three different values of $\ell$ and $m$.
This figure shows that for small $\ell$ few states $\nu'$ away from $\nu$ 
are significantly excited, resembling vertical transitions.
Thus, for small values of $\ell$, it might be reasonable to 
simplify expression \eqref{eq:pop2-1} by eliminating the sum and 
evaluating only the term with $\nu'=\nu$. 
However, for increasing values of $\ell$, more states with
different $\nu'$ contribute and the whole sum in Eq.\ \eqref{eq:pop2-1}
should be kept.
From this figure we see that there is also an $\ell$-dependent shift 
of the peak away from the value $\nu'=\nu$, which in our figure was 
chosen so as to cancel the $m$-induced shift.

Intersubband transitions are traditionally detected using relative
absorbance measurements in multi-quantum-well structures \cite{wes-egl}.
Normally, the transitions are produced by the transverse component
of the light's electric field at oblique incidence.
It is thus worth comparing the relative strength of those transitions to
our proposed normal-incidence transitions.
A simple estimate can be done by comparing the ratio of the longitudinal
to the perpendicular component of the vector potential given in 
Eq.\ \eqref{Eq_A}.
This ratio is given roughly by $q_r/q_z$, which can be chosen in the 
laboratory fairly freely.
If $q_r/q_z\approx 0.1$, then the population of the second subband states
is roughly a hundred times smaller than in the usual scheme.
Given the available laser powers this reduction does not represent a
significant experimental limitation. 
Alternatively, a pump-and-probe experimental scheme would also be adequate 
to observe the normal-incidence photoexcitation examined here.


\section{Conclusion}

We have briefly analyzed the viability of inducing intersubband
transitions in semiconductor quantum wells at normal incidence with
twisted light beams.
These transitions are enabled by the component of the electric field in 
the propagation direction which is present in collimated beams.
We obtained the matrix elements of the light-matter interaction, which display
the conservation law associated to the transfer of both spin and orbital 
angular momentum from the light to the electrons in doped quantum wells.
A simplified expression for the matrix elements, valid in the standard case 
of a somewhat small semiconductor system---as compared to the laser beam
waist---was presented and computed numerically.
Using the population of the second subband to second order in the field, 
we argued for the feasibility of the intersubband 
transitions induced by twisted light at normal incidence. 
 
\section*{Acknowledgments}

We acknowledge financial support from the Cooperation Program
ANPCyT--Max-Planck Society,
through grant PICT-2006-02134,
and from the University of Buenos Aires, through grants UBACYT 
2008/2010--X495 and 2011/2014--20020100100741.


\end{document}